\documentclass{ws-ijfcs}
\usepackage{enumerate}
\usepackage{url}
\begin{document}

\markboth{O. Cadenas and  G. M. Megson}
{Non-uniform quantization; linear average-case computation time}

%
\catchline{}{}{}{}{}
%

\title{Non-uniform quantization with linear average-case computation time}

\author{Oswaldo Cadenas}

\address{Division of Electrical and Electronic Engineering \\ London South Bank University, 103
Borough Road, London SE1 0AX, UK \\
\email{cadenaso@lsbu.ac.uk}}

\author{Graham M. Megson}

\address{School of Electronics and Computer Science\\
University of Westminster, London W1W 6XH, UK\\
g.megson@westminster.ac.uk}

\maketitle

\begin{history}
\received{(18 August 2021)}
\end{history}

\begin{abstract}
A new method for binning a set of $n$ data values into a set of $m$ bins
for the case where the bins are of different sizes is proposed. The method skips binning using 
a binary search across the bins all the time. It is proven the method exhibits a linear average-case computation time. 
The experiments' results show a speedup factor of over four compared to binning by binary search alone
for data values with unknown distributions. This result is consistent with the analysis of the method.
\end{abstract}


\section{Introduction}	

Binning data is about representing a large set of input data values (such as reals) into a smaller set of output values 
or bins. Binning is also often referred to as quantization \cite{1}. Each bin size is an interval given by the difference between a pair of neighboring bin boundaries; 
$m$ bins are defined by $m+1$ bin boundary values. Binning maps a data value $x$ into a bin as an integer $q$ in the range $1, \dots, m$. For example, time in hours is binned into 12 bins (or 24) with each bin 
of equal size of 60 minutes; this is \emph{uniform} quantization. Age of people binned into categories such as infant, 
child, teenager, etc. requires \emph{non-uniform} bins; that is bins of different sizes. Quantization is used everywhere in 
engineering; it is common in analog-to-digital conversions with uniform bins \cite{2} while non-uniform quantization is used for read-voltage level operations in multi-level NAND flash memory \cite{3}. 
Non-uniform quantization is used in audio coding standards and the most often used video encoding format and streaming video internet sources \cite{1} where modern embedded and mobile computer
systems are ubiquitous. Non-uniform bins result in methods 
to analyse the performance of embedded systems; e.g. centered bin distribution \cite{4}.
In data science, binning is frequently performed on continuous attributes. Here bin boundaries are adjusted by 
supervised methods on unknown data distributions leading to non-uniform bins \cite{5}, a method also used in cosmology research \cite{6}. 

Binning data values into uniform bins is straightforward and takes a little more than a simple division 
operation per data value. Binning into non-uniform bins requires more work. Linear search can be used across all the bin boundaries with a computational time cost of $O(m)$ per data value or more efficiently by performing a binary
search across the bin boundaries with a cost of $O(\lg m)$ computational time per data value \cite{7}. Recent research look at ways to define the size of the non-uniform bins for a particular problem \cite{6,8} rather than computing the 
quantization value across a set of non-uniform bins; this is what this paper addresses.     
Binary search is the  method used in functions for binning data in numeric computational packages such as Matlab, Python and R.  

This work proposes a binning method for a data value $x$ into $m$ non-uniform bins, $B$, which results in much faster binning than 
binary search. The method performs a one-off pre-computation at the outset. First, it forms $m$ new uniform bins, $U$, and then computes histograms of the $B$ bin boundaries 
within $U$. Each data value $x$
is binned in $U$ (taking constant computational time) and this integer result is combined with 
the pre-computed histogram to complete the calculation of the binning of $x$ in $B$. It is proven that this extra step requires a linear average 
time complexity producing significant speed-up to the binning process. 

\section{Proposed Binning Method}
\subsection{Binning}
Bins are defined as monotonically increasing boundary values $b_1, \dots, b_{m+1}$ such that $b_i < b_{i+1}$ for $i = 1, \dots m$ with $b_i \ge 0$. 
This defines $m$ bins. Binning a data value $x$ results in an integer value $q$, such that when
 $b_i \leq x < b_{i+1}$, then
$q = i$. Note that index bins start from 1. The binning process outputs integers $q$ in $1, \dots, m$ for data values within the bin boundaries; thus $q = 0$ means $x < b_1$ and $q = m+1$ means $x > b_{m+1}$ since it describes bin intervals that include the left 
boundary value. Therefore, the experiments in this paper use data values in the range $b_1 \leq x < b_{m+1}$. 
For non-uniform bins $b_{i+1} - b_i$ is not constant
for some $i$, and this set of non-uniform bins is referred to as $B$. 
In the uniform case the difference is constant for all bins, and the bins are referred to collectively as $U$. 
Binning $x$ in $B$ is denoted as $B(x)$. This work applies 
the binning process $B(x_j)$ to a data set $x_j$, for $j = 1, \dots, n$, with $n > m$, where $B$ stays fixed while $x_j$ 
might be constantly changing (e.g. streaming applications). 

\subsection{Outline of the new approach}

\begin{figure}[!t]
  \centering
  \includegraphics[width=120mm]{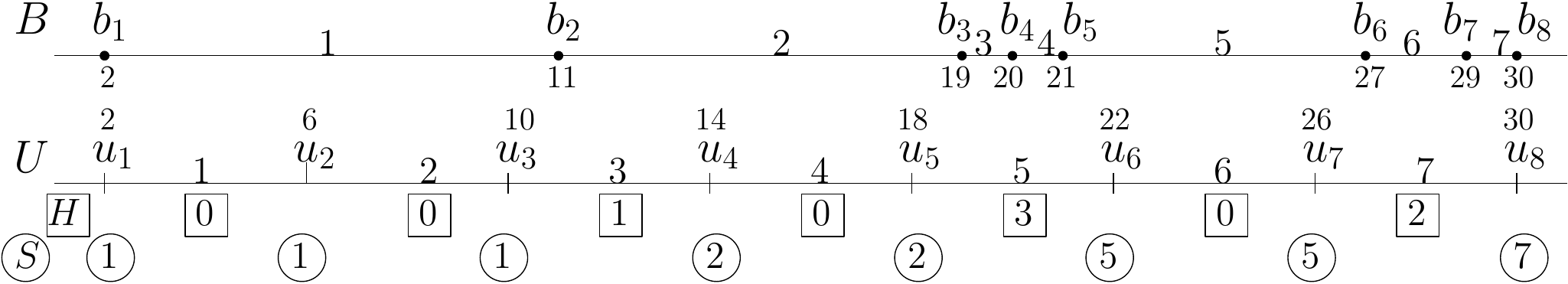}
  \caption{New binning supported by a histogram $H$ and a prefix sum $S$}
  \label{fig1a}
  \end{figure}

Consider bins $B$ with bin boundaries shown as dots on the upper line 
in Fig. \ref{fig1a}. Note all seven bin intervals are not of the same size. 
Any data value $x$ within these bins is mapped to an integer $q$
in the range $1, \dots, 7$; this is shown by the number in between the $b_i$ values 
(over the line in the figure). 

\emph{Step 1 - set up $U$:} Create $m = 7$ uniform bin intervals with size $(b_8 - b_1)/m$; these are shown as $U$ on the lower
line, also in Fig. \ref{fig1a}. The bin boundaries $u_i$ are shown at small vertical ticks. 
Note, the extreme boundaries of $B$ and $U$ are the same. 

\emph{Step 2 - histogram of $B$ in $U$:} Next compute the histogram of 
bin boundaries $b_2 \dots b_7$ as $H = h_1, \dots, h_7$ using $U$ as the bins for the histogram. These are the numbers
inside a square at the bottom of the figure. For instance, the bin that goes from $u_3$ to $u_4$
has $b_2$ in it, so a count of $h_3 = 1$ while the bin from $u_7$ to $u_8$ has $b_6, b_7$ in it, so a count of $h_7 = 2$.
Note that histogram $H$ is of length $m$; that is $H = h_1, \dots, h_m$ and that $b_1, b_{m+1}$ are excluded (since extreme values of $B$ and $U$ are the same) and so only $m-1$ boundaries of $B$ go into the histogram so that $\sum_i^mh_i = m-1$.   

\emph{Step 3 - cumulative histogram:} We then compute $s_{i+1} = s_i + h_i$ for $i = 1, \dots, m$ with $s_1 = 1$ (since $b_0 = u_0$, we do not bin $b_0$ in $U$ but assumed included in the histogram); this is the prefix sum of $H$ as $S$ or cumulative histogram of length $m+1$; $s_{m+1} = m$ since we started with $s_1=1$ and $\sum_i^mh_i = m-1$. The values $s_i$ are the
numbers in the circle at the bottom of the figure; they align with $u_i$. For instance, $s_6 = s_5 + h_5 = 2 + 3 = 5$.

The set up of $U$, $H$ and $S$ is performed once as a pre-computation before any data $x$ is binned. 
The binning of a data value $x$ starts by binning $x$ as $q = U(x)$ and then the method works out the bin value $B(x)$ from $q, H$ and $S$.  

\subsection{Examples of Binning}
Suppose $b_1, \dots, b_8$ of $B$ in Fig. \ref{fig1a} with values 2, 11, 19, 20, 21, 27, 29, and 30. Then $U$ is set up as 2, 6, 10, 14, 18, 22, 26 
and 30; with bin intervals of size 4 since $(30-2)/7 =  4$. The following examples illustrate binning to the correct $B$. 

\subsubsection{Case $h = 0$}
Consider binning value $x = 25$. Referring back to 
Fig. \ref{fig1a} we find $q = U(25) = 6$. The value $h_q = h_6 = 0$ so $B(25) = s_6 = 5$. From Fig. \ref{fig1a},
we see that $B(25) = 5$. A similar situation happens when $U(x) = 1, 2, 4$ in the figure. Notice that after pre-computation this case 
requires only a map to $U$ and a look-up. 
This case is computed fast. 

\subsubsection{Case $h = 1$}
Consider binning $x = 13$; $U(13) = 3$ and $h_3 = 1$. Look in $S$ to find $s_3 = 1 = r$; then we make the 
comparison $x \geq b_r$. In this instance $13 \geq 11$ and we calculate $B(13)$ as $s_{q+1} = s_4 = 2$; this is verified 
by Fig. \ref{fig1a}. Note that for $x = 10.5$, $U(10.5)$ is also 3 but $10.5 \geq 11$ is \emph{not true} and then $B(10.5)$
is calculated as $s_3 = 1$. With either comparison outcome this case mapping is simple and fast.  

\subsubsection{Case $h = 2$}
When binning an $x$ value such that $h_q = 2$ for $q = U(x)$, 
proceed similarly to discover that two comparisons are needed to calculate $B(x)$. For instance when $U(x) = 7$ in the above example. This case requires constant time.

\subsubsection{Case $h > 2$}
This case occurs when $h_q > 2$; $U(x) = 5$ in the example. In this case, we extract the bin boundaries $b_3, b_4, b_5$ from $B$ to form a
smaller set of bins ($h_q-1$ bins); a binary search within this smaller set of bins gives the information required to determine $B(x)$. 
For instance, if $x = 19.5$, $q = U(19.5) = 5$, $h_q = h_5 = 3$. We obtain $r = s_q = s_5 = 2$. A binary search, $\verb"BSearch"(x, [b_3, b_4, b_5])$ 
returns 1 and so $B(19.5) = r + 1 = 2 +1 = 3$ as expected from looking at Fig. \ref{fig1a}. The surprising result of this paper is that $h_q < 4$ for the average case and so this case again is performed in constant time on average.

\section{Algorithm and Model}

\subsection{Algorithm}
Processing data with the proposed method follows straightforwardly from the discussion above. We assume $B$ is given as $b_1, \dots, b_{m+1}$ 
as previously defined. We are binning input $x_j$ for $j = 1, \dots, n$ with values in any order and with any 
statistics. The output is $q_{x_j} = B(x_j)$. 
\begin{enumerate}
\item Step 1: Set up bins $U$ as $u_1, \dots, u_{m+1}$ as $u_{i+1} = u_i + \Delta$ for 
$i = 1, \dots m$ and $\Delta = (b_{m+1} - b_1)/m$ with $u_1 = b_1$. Note $u_{m+1} = b_{m+1}$.  
\item Step 2: Compute the histogram $H = h_1, \dots, m$ as $h_{q} = h_{q} + 1$ with $q = U(b_i)$ for all $b_i$ of $B$ in $i = 2, \dots, m$.  
\item Step 3: Compute the prefix sum of $H$ as $S$, of length $m+1$, with $s_1 = 1$ and $s_{i+1} = s_{i} + h_{i}$ for $i = 1, \dots, m$.
\item Step 4: For each $x_j$ in $j = 1, \dots, n$ obtain $q = U(x_j)$, $h = h_q$ and $r = s_q$, then do either one of:
\begin{enumerate}
  \item Case $ h = 0$: $B(x_j) = r$.
  \item Case $ h = 1$: if $x \geq b_{r+1}$ then $B(x_j) = r+1$ else $B(x_j) = r$. 
  \item Case $ h = 2$: if $x \geq b_{r+2}$ then $B(x_j) = r+2$ \\
  \hspace*{1.4cm} else if $x < b_{r+1}$ then $B(x_j) = r$ \\
  \hspace*{1.4cm} else $B(x_j) = r+1$.
  \item Case $ h > 2$: $B(x_j) = r + \verb"BSearch"(x, B_q)$ \\
  \hspace*{1.4cm } $B_q$ is the subset of $b_i$ where $U(b_i) = q$. 
\end{enumerate}
\end{enumerate}

Steps 1, 2 and 3 are performed once and so its computational time cost
diminishes quickly for large $n$. $B_q$ can be pre-computed as part of Step 2 and necessary only for those instances where the histogram count is greater than two.
This is left down to a programming implementation of the method. Cases $h = 0, h = 1$ and $h = 2$ require 0, 1 and 2 comparisons respectively and thus are of
constant computation time per $x$ data item. The case $h > 2$ is considered in the next section. 

\subsection{Setting Uniform Bins U} As boundary values of $B$ are defined such that $b_i < b_{i+1}$ then $B$ has $m$ intervals greater than zero since $b_{i+1} - b_i > 0$. The range of values $b_{m+1} - b_1$ can be divided into $m$ intervals of equal size with boundary values $u_1, u_1+\Delta, \dots, u_1+m\Delta$ with $u_1 = b_1$ and $u_1+m\Delta = b_{m+1}$ for $\Delta = (b_{m+1}-b_1)/m$. These are the uniform bins of $U$ with the span of $B$. Binning a value $x$ in $U$, $U(x)$, takes constant time by mapping $x$ into integer $q =1, \dots, m$ as $q = \lfloor\frac{x-u_1}{\Delta}\rfloor+1$.
\subsection{Histogram of $U(b_i)$} Note $U(b_1) = 1$ and $U(b_{m+1}) = m+1$ with $b_{m+1}$ being the smaller value that is mapped to $m$ since $b_{m+1} > b_m$. This implies that $U(b_i)$ for $i = 2, \dots, m$ may map into any integer $q$ in the range $1, \dots, m$ but as there are $m$ bins in $U$ and we are only mapping $m-1$ $b_i$ boundaries values then, the pigeon hole principle conditions are not met ($m-1$ pigeons, $m$ holes) to guarantee that at least one $b_i$ boundary value of $B$ will be found within a $U$ interval (see Fig. \ref{fig1a} where $b_2 - b_1 > \Delta$ for example). Thus, defining $h_q$ as the count of cases where $U(b_i) = q$ for $i = 2, \dots, m$ we can guarantee that $\sum_{q=1}^m h_q = m-1$; this is computed and stored over $U$. 

\subsection{Cumulative Histogram of $U(b_i)$} We then compute the cumulative histogram $S$ as $s_{i+1} = s_i + h_i$ for $i = 1, \dots, m$ with $s_1 = 1$ (to include $b_1$) so that $s_{m+1} = m$. Note that $s_i$ are  annotated aligned with $u_i$ boundary values. 
We reason about $s_i$ in these terms. For a value $x$ such that $U(x) = q$ we look at the value $r = s_q$ and that means that the $x$ value is greater or equal to the first $r$ boundary values of $B$, that is $x \ge b_i$ for $i = 1, \dots, r$. Note that $q = B(x)$ also means that  $x$ lies within the uniform interval $q$ that goes from $u_q$ to $u_{q+1}$ boundary values of $U$, and so to obtain the actual binning $B(x)$ we have to resolve where $x$ actually lies within the possible $b_i$ boundary values that also mapped as $q = U(b_i)$; information already available in $h_q$.

\subsection{Computing $B(x)$} From above, for a value $x$ such that $U(x) = q$ we look at the value $r = s_q$. When $h_q = 0$, this means that the $q$ interval of $U$ does not contain any $b_i$ boundary value within and as such $B(x) = r$ by definition. No extra comparison is needed. 
When $h_q = 1$, this means that the $q$ interval of $U$ has one $b_i$ boundary value within it. This boundary value is $b_{r+1}$ and one comparison is needed in this case to determine $B(x)$ as either $r$ or $r+1$.  
When $h_q = 2$, this means that the $q$ interval of $U$ has two $b_i$ boundary values within it. These are boundary values $b_{r+1}, b_{r+2}$ and so we need two comparisons for this case to determine $B(x)$ as either of $r$, $r+1$,  $r+2$.
When $h_q > 2$, then we do a binary search of $x$ within the subset of boundary values $b_{r+1}, b_{r+2}, \dots, b_{r+h_q}$ to find the offset to be added to $r$ to determine $B(x)$. 

\section{Analysis}
\subsection{Average Value of cases $h > 2$} The binning problem reduces to how we count all the possible valid solutions of distributing $m_1$ boundaries ($B$) across $m_2$ ($U$) bins. 
Define the mappings of non-uniform boundaries to uniform bins by a $C \times m_2$ 2D grid of values $0 \leq j \leq m_1$. 
The 2D grid $m_2C$ amounts to the global count of bins with valid $h_i$ values. 
For the whole grid, the amount of bins having a value of $j$ is proportional to $C_j$.

As a simple example consider $m_1 = m_2 = 3$. All possible solutions where 3 bins add up to 3 are: 300, 210, 201, 120, 102, 111, 030, 021, 012, and 003. 
All possible solutions can be expressed in a $10 \times 3$ grid, $C = 10$; 10 rows of 3 slots each. The 30 slots in the grid satisfy $3(10) = 3(4+3+2+1) = 12 + 9 + 6 + 3$; there are 12 slots with a count of 0, 9 slots with a count of 1, 6 slots with a count of 2 and 3 slots with a count of 3. The $C_j$ counts are the 4, 3, 2, 1 values above where $C_0 = 4$. General expressions are given below.   

\subsubsection{Value of $C$} Each row of the grid satisfies the equation $\sum_{j=1}^{m_2} h_j = m_1$ then by the stars and bars approach 
we see that there are $C = {{m_1 + m_2 -1} \choose {m_1}}$ rows to the grid \cite{12}.

\subsubsection{Value of $C_j$} When the value of a slot in a row on the grid  is $j$ then $m_2-1$ slots add up to $m_1-j$ which implies there are 
${{m_2-1 + m1 - j -1} \choose {m_1-j}}$ such slots for any value of $j$ in $0, \dots, m_1$ then $C_j = {{m_2 + m1 - (j+2)} \choose {m_1-j}}$.

\subsubsection{Counting the grid} Equation,
\begin{equation}
m_2C = m_2\sum_{j=0}^{m_1}C_j  
\label{eq:eq1}
\end{equation}
follows by a counting rule on the grid.

Also note that the sum of all possible values for the $C$ solutions is,
\begin{equation}
m_1C = m_2 \sum_{j=0}^{m_1}jC_j
\label{eq:eq2}
\end{equation}

Dividing equation Eq.~(2) by equation Eq.~(1) gives an average of the values of a slot in the grid as

\begin{equation}
\frac{m_1C}{m_2C} = \frac{m_2 \sum_{j=0}^{m_1}jC_j}{m_2\sum_{j=0}^{m_1}C_j} = \mu_{all} = \frac{m_1}{m_2}
\label{eq:eq3}
\end{equation}

Defining $P_j = C_j/C$ allows writing this relation as,

\begin{equation}
\begin{aligned}
\frac{m_1}{m_2} = \frac{m_2 \sum_{j=0}^{m_1}jC_j}{C} \\
\frac{m_1}{m_2} = \sum_{j=0}^{m_1}jP_j = \mu_{all}
\end{aligned}
\label{eq:eq4}
\end{equation}
Note that $\sum_{j=0}^{m_1}P_j = 1$; $0 < P_j < 1$ are Poisson-like probability distribution for the slot values $j$. 

We can derive a recurrence,
 
\begin{equation}
\begin{aligned}
P_{j+1} = \frac{(m_1-j)}{(m_2+m_1-(j+2))}P_j \\
P_0 = \frac{(m_2-1)}{(m_1+m_2-1)} 
\end{aligned}
\label{eq:eq5}
\end{equation}
where $P_{j+1} < P_j$.

\subsubsection{Average value for $\mu_{>2}$} From Eq.~(4), average 
\label{equations}
\begin{equation}
\mu_{>2} = \frac{\sum_{j=3}^{m_1}jP_j}{\sum_{j=3}^{m_1}P_j} = 3 + \frac{\sum_{j=1}^{m_1-3}jP_{j+3}}{P}
\label{eq:eq6}
\end{equation}

with $P=\sum_{j=3}^{m_1}P_j$. 
We know $P_1 + 2P_2 + \sum_{j=3}^{m_1}jP_j = \mu_{all}$  and also $P_0 + P_1 + P_2 + P = 1$.
Therefore,
\begin{equation}
\mu_{>2} = \frac{\mu_{all}-P_1-2P_2}{1-P_0-P_1-P_2} 
\end{equation}
As $\sum_{j=1}^{m_1-3}jP_{j+3} < \mu_{all}$ then it follows,

\begin{equation}
3 < \mu_{>2} < 3 + \frac{1}{1-P_0-P_1-P_2}\left( \frac{m_1}{m_2} \right) 
\end{equation}
When $m = m_1 = m_2$ this implies $\mu_{>2} < 4$ for large $m$. Also as $m_2 \to \infty, \mu_{>2} \to 3$.

Back to the small example with $m_1 = m_2 = 3$, from Eq.~(5) $P_0 = 0.4, P_1 = 0.3, P_2 = 0.2$ and hence $P = 0.1$. 
So, P = 0.1 implies that 10\% of the 30 slots in the grid are slots with counts greater than 2, that is 3 in this case as seen previously.  
Using the recurrence from Eq.~(5), and for $m_1 = m_2 = m$ we have that,

\begin{equation}
\lim_{m \to \infty} P_0 =\frac{1}{2}; \lim_{m \to \infty} P_1 =\frac{1}{4}; \lim_{m \to \infty} P_2 =\frac{1}{8}
\end{equation}

This is a hint that bins with counts greater than 2 are of around 12.5\% ($1-P_0-P1-P_2$) and from Eq.~(8) the count is capped to 4 on average. 

\subsection{Time Complexity} Let $\mu$ be the maximum number of $b_i$ from $B$ mapped to a single bin in $U$. The complexity of the method is
$t(n) \approx a_0(n) + 2a_1(n) + 3a_2(n) + \sum_{i=3}^{q}a_{i}(n)(1+\lg{\mu})$ where $a_i(n)$ is the count of items processed under the four cases 
determining $B(x)$. Gather terms together and simplify by putting $t(n) \leq 3a_{012}(n) + a_{q>2}(n)(1+\lg{\mu})$ where $a_{012}(n)= a_0(n) + a_1(n) + a_2(n)$ and 
$a_{q>2} = \sum_{i=3}^q a_i(n)$. This makes use of the fact that $\mu$ is a constant for a particular run of data. As there are $n+1$ ways to 
distribute the computations the average workload is given by:
\begin{equation}
t_{avg}(n) \leq \frac{1}{n+1}\left[3 \sum_{i=0}^na_{012}(n-i) + \sum_{i=0}^n a_{q>2}(i)(1+ \lg{\mu})\right]
\end{equation}
We note that $a_{012}(n-i) = n- i$ and $a_{q>2}(i) = i$ since it does not matter how the data items are arranged with the bins associated with each of the 
aggregated terms. Thus, 
\begin{equation}
\begin{gathered}
t_{avg}(n) \leq \frac{1}{n+1}\left[3 \sum_{i=0}^n(n-i) + \sum_{i=0}^n i(1+ \lg{\mu})\right] \\
t_{avg}(n) \leq \left[ 2n + \frac{n}{2} \lg {\mu}\right]
\end{gathered}
\end{equation}
 which implies $t_{avg}(n) \leq O(n \lg{\mu})$. And since $\mu < 4$ we conclude the 
method has $O(n)$ time average performance.  

\section{Experiments and results} 
\begin{figure}[!t]
\centerline{\includegraphics[width=6cm]{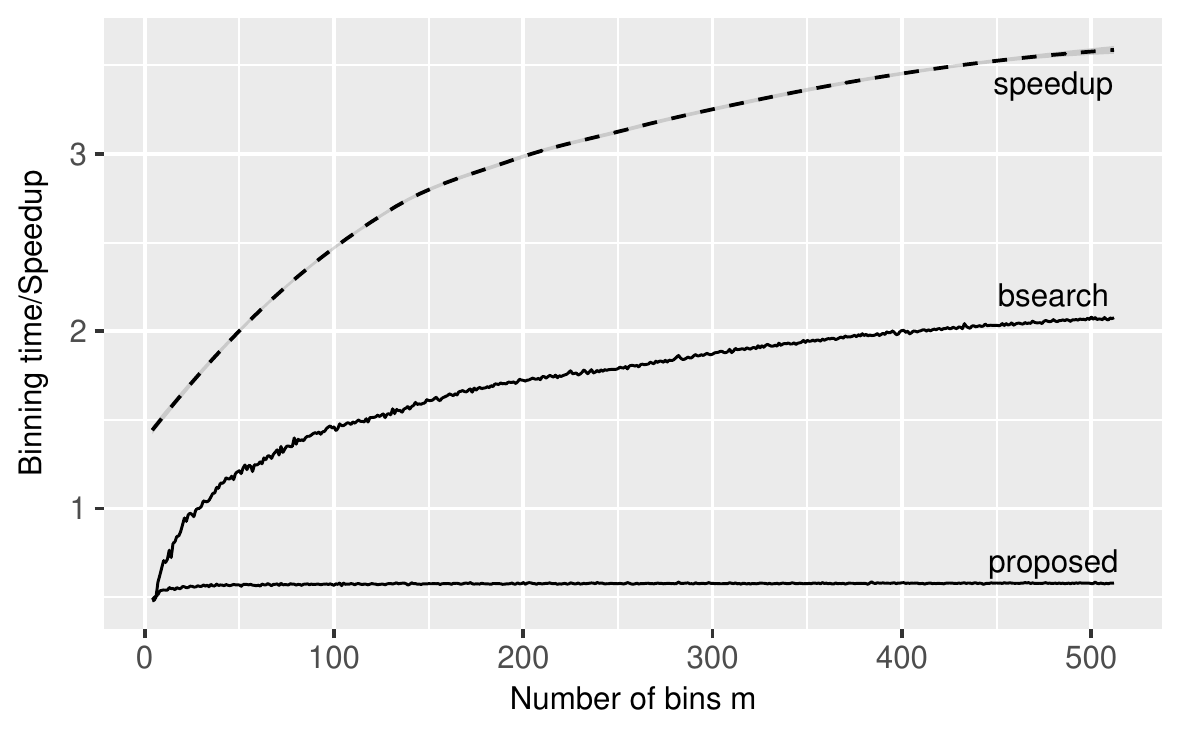}}
\caption{Binning performance using the proposed method and binary search. Binary search (middle line) shows a $O(n \lg m)$ behaviour as expected; proposed binning method (lower line) is faster than binary search over all of $m$.}
\label{fig2}
\end{figure}

Two main experiments were conducted on an x86-64 microprocessor. Firstly, binning data within a random set of $m$ bins, $B$, with 
$3 < m \leq 512$. Twenty million $x_j$ values were used with different random distributions. Five thousands of runs were
performed to compute $B(x_j)$, both with a binary search and with the proposed binning method and then average taken. Fig.~\ref{fig2} shows a faster 
computation with the proposed method, over
binary search, as the number of bins increases ($m_1 = m_2 = m$). The speedup factor is well over a factor of three 
for large $m$. This is so as binary search time follows a $O(n \lg m)$ time as expected, while the proposed computation method 
exhibits time $O(n)$ on average. 

\begin{figure}[!t]
\centerline{\includegraphics[width=60mm]{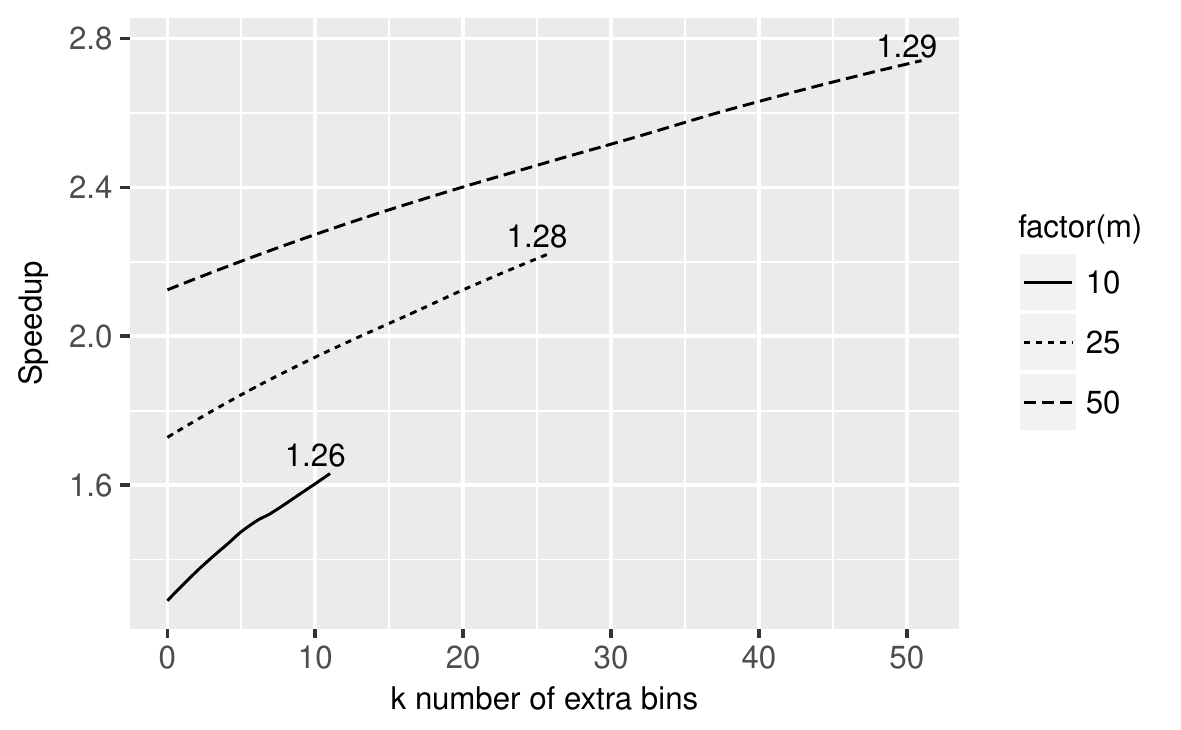}}
\caption{Extra computation speedup using $m+k$ bins when setting up $U$. An extra speedup factor of around 1.25 is achieved for $k=m+1$.}
\label{fig3}
\end{figure}

In a second experiment, the data set $x_j$ is kept fixed. Binning by binary search is performed for $m = 10, 25, 50$ number of bins; these are typical values for binning when exploring data. Binning with the proposed method is then performed on the same 
$x_j$ set using $m+k$ bins when setting up $U$; parameter $k$ is varied across the range $k = 0, \dots, m+1$ over thousands of runs. 
The results, shown in Fig.~\ref{fig3}, show that an extra speedup factor is gained for $k > 1$ and reaches around a factor of 1.25 
when $k = m+1$. This simple extra arrangement results in an overall combined speedup factor of over four by essentially having $U$ with double the size of bins of $B$. 

\section{Discussion}
The proposed method for binning a value $x$ into $m$ non-uniform bins decomposes the process 
into four base cases, three of which take constant computation time. A fourth 
case, recurs to binary search but only of $h$ bins with $h < m$; $3 < h < 4$ on average as shown. Other methods to improve over binary search, 
such as interpolation search, that has 
a $O (\lg \lg m)$ time per item, only benefits uniformly distributed data \cite{9}. We did not find a 
substantial benefit of interpolation search over binary search in the two experiments conducted here. The proposed method is expected to improve 
processing techniques that use binary search such as fractional cascading \cite{10}
that searches over multiple sequences or for finding interval intersections in gene sequencing \cite{11}. The analysis shows that for large values of $m$ the probability 
of having bin count values greater than 2 is of 12.5\% and of 87.5\% of having a bin count of either 0, 1 or 2. 
A closed formula of the average of any bin having a count greater than 2 is $\frac{m_1 -m_2(P_1+2P_2)}{m_2(1-P_0-P_1-P_2)}$, for any $m_1, m_2$ (see section~\ref{equations}). By doubling the bins in $U$ compared to $B$ the 
probability of a bin count being 0, 1 or 2 increases to over 95\%. Indeed the probabilities, when doubling the bins in $U$ ($m_2 = 2 m_1$ and for large values) are
$P_0 = 2/3, P_1 = 2/9; P_2 = 2/27$ for a bin count of 0, 1, and 2 respectively with a bin count average of $7/2$ when the method recurs to binary search. 
With binary search, processing $n$ data items requires
$t_{bs} = n \lg m$ time. With the proposed method, it requires $t_p = 1nP_0 + 2nP_1 + 3nP_2 + (1+\lg \mu) (1-P_0-P_1-P2)n$ which reduces to 
$t_p = 1.75n$ given an theoretical speedup of $\frac{t_{bs}}{t_p} \leq 0.6 \lg m_2$. This translates to an speedup of around 5 for $m=512$ which is consistent 
with the speedup shown in Fig. \ref{fig2} and consistent with the analysis in Section 4. By doubling the bins in $U$, the speedup increases by a factor of around 1.26 also consistent with the results. 

\section{Conclusion}
It was proven that on average the binning method presented here, for non-uniform quantization, runs in linear time and so it is faster for binning the same data values 
when compared to using binary search. The method gets better as $m$ increases. Empirically, we find a speedup factor of over three 
for large $m$; result explained from the theoretical analysis. Extra acceleration 
is achieved by adding a parameter $k$, $k=m+1$ (essentially doubling the number of bins in the internal mechanism used in the method), we report an extra speedup factor of 1.25 which is also consistent with the analysis. This parameter $k$ increases the probability of binning a value 
within $m$ non-uniform bins in constant computational time. 
The method applies equally to real or integer values and it is directly 
applicable to computing histograms. Results shows the binning of data with the proposed method performs four times faster than existing methods in streaming applications using standard microprocessors.


\end{document}